\documentclass[prl,twocolumn,showpacs,amsmath,amssymb]{revtex4}

\input epsf
\usepackage{graphicx}
\usepackage{dcolumn}
\usepackage{bm}

\begin{document}

\preprint{}

\title{Contact dynamics in a gently vibrated granular pile}

\author{Alexandre Kabla}
\author{Georges Debr\'egeas}
\email{georges.debregeas@college-de-france.fr}
 \affiliation{Coll\`ege de France, France\\}
\date{\today}

\begin{abstract}

We use multi-speckle diffusive wave spectroscopy (MSDWS) to probe the
micron-scale dynamics of a granular pile submitted to discrete
gentle taps. The typical time-scale between plastic events is
found to increase dramatically with the number of applied taps.
Furthermore, this microscopic dynamics weakly depends on the solid
fraction of the sample. This process is strongly analogous to the
aging phenomenon observed in thermal glassy systems. We propose a
heuristic model where this slowing down mechanism is associated
with a slow evolution of the distribution of the contact forces
between particles. This model accounts for the main features of
the observed dynamics.

\end{abstract}

\pacs{81.05.Rm, 45.70.Cc, 05.45.-a}

\maketitle

Internal contact forces in dense granular systems are very
inhomogeneous \cite{Mueth1998,Lovoll1999}, even for crystalline
assemblies \cite{Blair2001}. The stress at a given contact can
change macroscopically following a relative displacement of the
two particles of order of microns. Hence, large modifications of
the contact forces field can result from minute deformations of
the pile. This phenomenon is crucial in understanding the
catastrophic yielding occurring in granular systems submitted to a
slowly varying stress (avalanches \cite{Staron2002}, shear-bands
in triaxial tests \cite{Desrues1996}). It also explains why the
sound transmission through a granular sample can be strongly
affected by very small deformations \cite{Liu1994}. To probe the
evolution of the internal stress field, one needs to measure
forces directly \cite{Howell1999, DaSilva2000} which is difficult
in 3D . In this letter, we propose a different approach: we use
MSDWS to measure particle displacements on micron-scales in a pile
submitted to gentle discrete taps. These vibrations are too weak
to induce large-scale rearrangements which would eventually lead
to a compaction of the granular system
\cite{Knight1995,Philippe2002,Lesaffre2000}. We can therefore
evaluate the micro-dynamics of the contacts without
significatively perturbing the packing structure.

We use glass beads of diameter $45\pm2\;\mu$m, contained in a
glass cell (30 mm $\times$ 10 mm $\times$ 2 mm). To reduce
electrostatic forces and the effects of moisture, the granular
system is saturated with pure water. During the experiment, the
mean packing fraction $\phi$ of is obtained by measuring the
position of the upper surface of the pile with a CCD camera.
Although a systematic error of 2\% can not be avoided, this allows
us to detect relative changes in $\phi$
 as small as $0.01\%$. To produce motion in the
pile, we use a piezoelectric actuator on which the cell is rigidly
mounted. Vertical vibrations of precisely controlled amplitude,
shape and durations can thus be applied to the granular column. In
this experiment, we focus on a single type of mechanical
excitation, later referred to as a "tap", which consists in a
train of square wave vibrations of frequency 1 kHz and duration
100 ms. Different applied voltage are used, yielding various
vertical amplitudes ranging from 50 to 300 nm.

In a standard experimental run, the pile is prepared by turning
the cell upside down then allowing the particles to sediment for
half an hour. This procedure yields reproducible structures of low
volume fraction. The pile is then submitted to high amplitude taps
(of vertical amplitude 300 nm) until it reaches a prescribed
packing fraction $\phi_{s}$. During this compaction stage, the
evolution of the packing fraction $\phi$ with the number of taps
(Fig. 1) is highly reproducible, and consistent with previous
experimental results on dry granular systems. We however note that
the packing fraction we reach is well above the close packing
limit expected for a fully disordered pile ($\simeq 0.64$). This
indicates that cristallisation does occur in our system under
vibration. This first step is for us a mean to reproducibly
prepare a granular sample of given packing fraction with
essentially the same preparation history. We then start probing
the dynamics of contacts by submitting the cell to very gentle
taps (of amplitude 50 nm). As shown in Fig. 1, this second step
does not induce significative evolution of the packing fraction
except for initially very loose packs.

\begin{figure}[!h]
\centerline{ \epsfxsize=8.5truecm \epsfbox{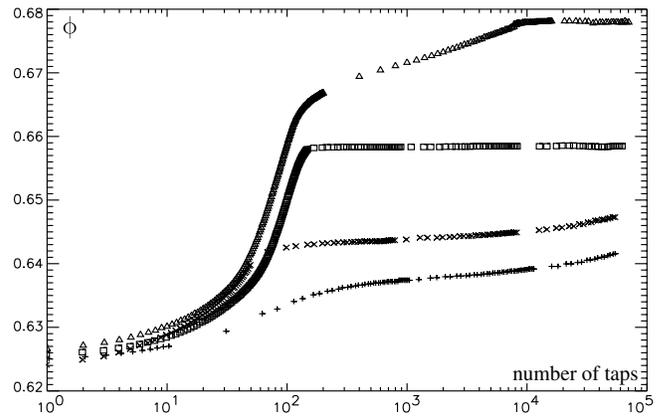}}
 \caption{Four compaction curves : $(+)$ is only excited by gentle vibrations;
The other systems are prepared with $(\times)$ 50, $(\Box)$ 150,
$(\bigtriangleup)$ 8000 high impulsions pulses, before being
excited by gentle vibrations. There is a systematic error of 2\%
on the measurements of the packing fraction.}
 \label{figure1}
 \end{figure}

To probe the microscopic dynamics induced by these gentle taps, we
use MSDWS \cite{Cipelletti1999,Viasnoff2002}. This technique,
which allows one to resolve sub-micron displacements, has been
successfully applied to granular dynamics by several groups
\cite{Menon1997,Kim2002}. The sample is illuminated with a He-Ne
laser beam at a depth of 2 cm below the pack upper surface (1 cm
over the bottom). Photons are multiply scattered by the particles
\cite{note-l-star}, and form a speckle pattern on the opposite
cell wall that we record with a CCD camera. In the absence of
vibrations, the speckle image does not change in time as
temperature is insignificant for such large objects. By contrast,
the taps induce some irreversible particles displacements which
modify the speckle image. To quantify this internal dynamics, we
measure the intensity correlation of speckle images, taken between
taps, as a function of the number of taps $t$ that separate them.
This function generally depends  on the total number of small
amplitude taps $t_w$ that have been performed. We therefore
calculate the two-times correlation fonction $g(t_w,t)$:

\begin{equation}
g(t_w,t) = \frac{{ \langle I(t_w  + t) \cdot I(t_w ) \rangle
_{spkl} - \langle I(t_w ) \rangle _{spkl}^2 }} {{ \langle I(t_w
)^2  \rangle _{spkl}^{} - \langle I(t_w ) \rangle _{spkl}^2 }}
\label{eqcorrel}
\end{equation}

\noindent In this expression, $\langle \; \rangle _{spkl}$
corresponds to averaging over different speckles. MSDWS thus
allows one to rapidly access relaxation time by substituting time-
with space-averaging and is therefore well suited to the study of
non-stationary dynamical systems.

\begin{figure}[!h]
\centerline{ \epsfxsize=8.5truecm \epsfbox{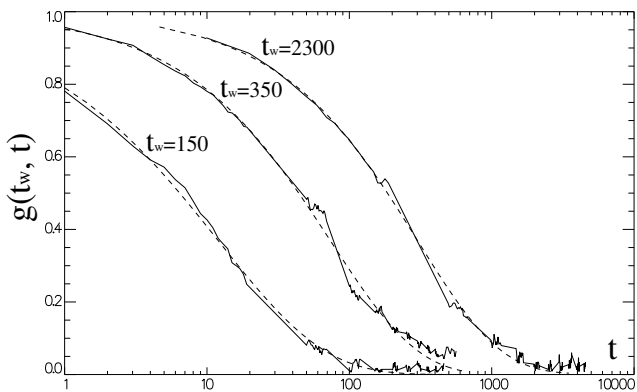}}
\caption{Three different correlation fonctions obtained after
three different times $t_w$. Solid lines correspond to
experimental values and dashed lines to the stretched exponential
fit.}
 \label{figure2}
 \end{figure}

Figure \ref{figure2} shows three correlation fonctions obtained
with the same sandpile at different values $t_w$. These fonctions
are well fitted by stretched exponentials:

\noindent $ g(t_w,t) = \exp \left( - \left( t/\tau(t_w)
\right)^{\alpha(t_w)}\right) $

\begin{figure}[!h]

\centerline{\epsfxsize=8.5truecm \epsfbox{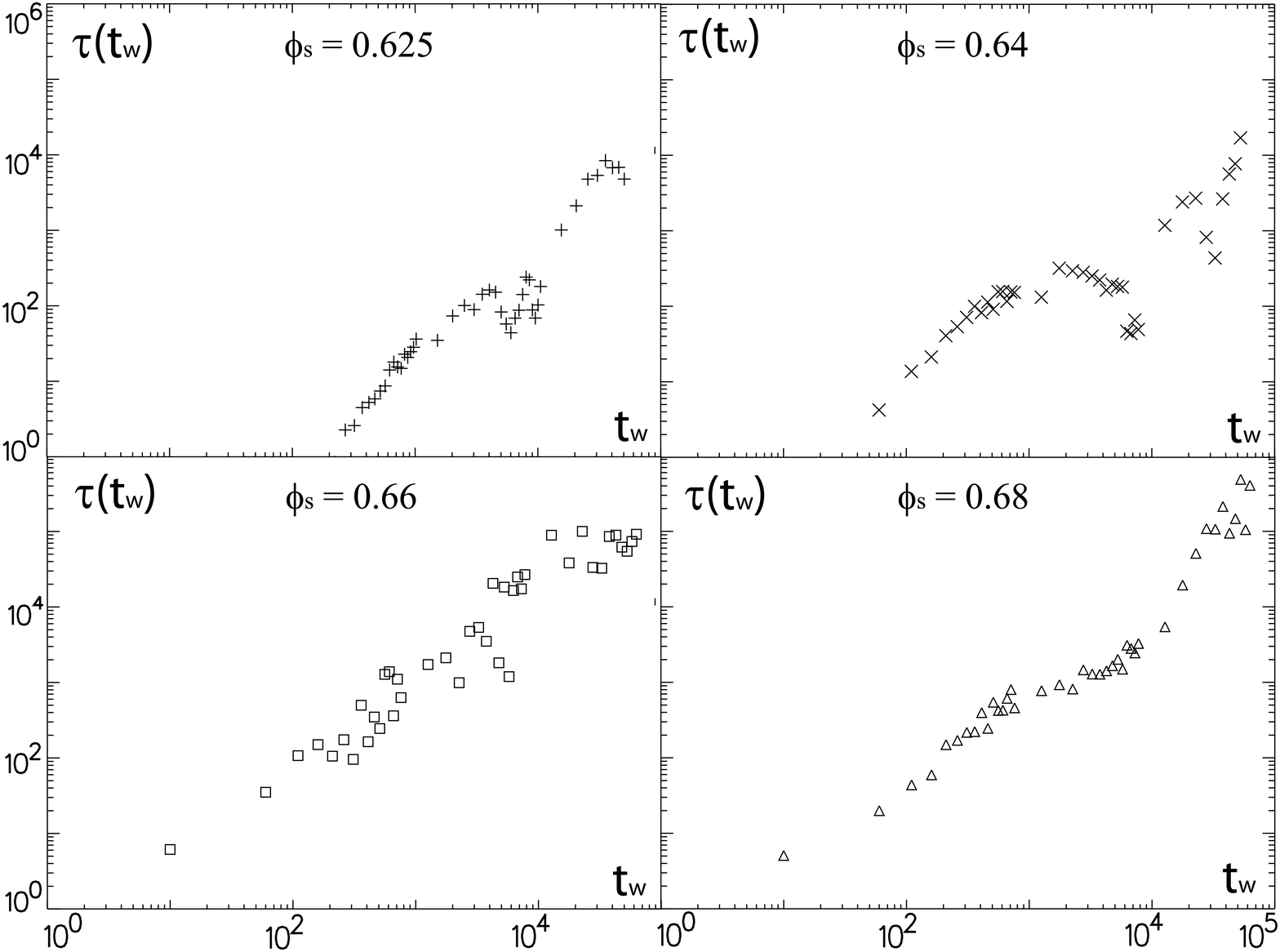}}
\caption{Evolution of the dynamical time $\tau$ with the number of
low magnitude taps $t_w$ for different packing fractions $\phi_s$}
 \label{figure3}
 \end{figure}

For different packing fractions, we follow the evolution of the
dynamics by monitoring the two parameters $\tau(t_w)$ and
$\alpha(t_w)$ as a function of the total number of gentle pulses
$t_w$. We find that the exponent $\alpha$ is roughly constant
($\simeq 0.8 \pm 0.2$) and independent of the packing fraction $\phi_s$.
By contrast, the time
$\tau(t_w)$ increases by five decades over the range of $t_w$
explored, as shown in figure \ref{figure3}. It should be noted
that this dynamics can be immediately reset by submitting the
system to a few taps of larger intensity (such as those used for
compacting the sample). A careful examination of the $\tau(t_w)$
curve also reveals large fluctuations in the internal dynamics,
especially in looser packs for which the packing fraction slowly
evolves. During certain periods of time, the dynamics is restarted
as shown by a sudden decrease of $\tau(t_w)$. This may correspond
to catastrophic failures of the pack structure, which compete
against a global reinforcement of the granular contacts.\\

This result demonstrates that the response of a granular system to
small perturbations is strongly dependent on the history of its
preparation (the number of applied taps $t_w$), and rather
insensitive to the packing fraction. More quantitatively, the
typical relaxation time associated with a given mechanical
excitation evolves as a power law of the total number of taps
$t_w$ : $\tau(t_w) \sim t_w^{1.2 \pm 0.2}$. This observation is
strongly reminiscent of the aging behavior recently observed in
many glassy colloidal systems
\cite{Knaebel2000,ViasnoffPRL2002,Cipelletti2003}. In these
materials, the longest ($\alpha-$)relaxation time is found to grow
as a power law of the time since the system was left to rest after
a rapid shearing. In spite of this strongly analogous behavior,
the microscopic processes leading to this dynamical arrest are
qualitatively different. In colloidal systems, stress relaxation
occurs by thermally activated rearrangements of the structure. In
granular materials, temperature is effectively zero and relaxation
only results from externally applied vibrations which induce the
slippage of some contacts (the most fragile ones).

The frequency of these slipping events is directly probed by the
MSDWS technique. Slipping events indeed result in some grain
displacements (either translations or rotations) around the broken
contact, until a new equilibrated configuration is found. In this
gentle vibrations regime, these events are too small to contribute
in a significant way to the compaction process. This means that
the associated grain displacements are of order $\delta << D$,
where $D=45 \; \mu$m is the particle diameter.  Although tiny,
these irreversible displacements control the speckle
decorrelation. However, to get a quantitative estimate of their
frequency, one needs to assume that they are uniformly distributed
in space and have a unique characteristic amplitude $\delta$.
Within this hypothesis, $\tau(t_w)$ is directly proportional to
the inverse of the yielding frequency \cite{DWS,Menon1997}.\\

We now turn to a tentative microscopic model to capture this
slowing down process. As underlined before, the observed dynamical
arrest is rather insensitive to the packing fraction of the
sample. We thus need to introduce another internal variable that would
 control the instantaneous response of the
pack to gentle vibrations. Here we propose to focus on contact
stress distribution. It has been observed that the form of this
distribution is almost independent of the volume fraction of the
pack and the preparation history. However, standard measurements
are not sensitive enough to detect small variations in these
distributions especially in the low force limit, that may follow
from very gentle mechanical vibrations. We will argue here that
the observed evolution of the dynamics results precisely from a
slow modification of the stress distribution inside the pack which
effectively rigidify the granular pile.

To modelize such a dynamics, we picture the granular assembly as a
set of independent contacts (which total number is supposed to be
a constant.) Each contact is characterized at time $t$ by the
normal and tangential component of the contact force which we
denote $\sigma_n$ and $\sigma_t$ respectively. Mechanical
equilibrium imposes that $\sigma_t<\mu \sigma_n$. For simplicity,
we will make the friction coefficient $\mu$ equal to $1$ in the
rest of the letter. For a given packing structure, the state of
the internal stress field is characterized by the two variable
stress density distribution $P_{\sigma}(\sigma_n, \sigma_t)$. As
the cell is vibrated, mechanical waves travelling through the
sample induce random stress fluctuations on each contact. Such
perturbations can locally trigger the rupture of a contact,
whenever the shear force $\sigma_t$ overcomes the normal force
$\sigma_n$. We assume an exponential distribution $\chi(\delta
\sigma)$ of the maximum force fluctuation induced by the tap on
each contact:

\begin{equation}
\chi(\delta \sigma) = \frac{1}{\overline{\delta\sigma}} \exp
\left( {- \frac{\delta \sigma}{\overline{\delta\sigma}}} \right)
\label{stressfluct}
\end{equation}

In this expression, the mean force fluctuation
$\overline{\delta\sigma}$ is an increasing function of the applied
vibration amplitude. Thus the probability for a given contact to
yield following a single tap writes:

\begin{equation}
\omega_y(\sigma_n, \sigma_t) = \exp \left( {- \frac{\sigma_n -
\sigma_t}{\overline{\delta \sigma}}} \right)
\end{equation}

This expression is a consequence of the peculiar form (Eq.
\ref{stressfluct}) taken for the tap induced force fluctuations
$\chi(\delta \sigma)$. However, the main results of the present
model remain valid for any fast decaying distribution (faster than
a power law).

After a yielding event, the force at the renewed contact is chosen
from a given "rejuvenated" distribution which we consider
intrinsic to the system. The distribution of normal forces in a
granular pile under moderate load is known to exhibit an
exponential tail at high forces and a plateau below the mean force
\cite{Mueth1998,Lovoll1999,Blair2001}. Numerical measurements have
also shown that, for a given value of the normal force, the
tangential forces are uniformly distributed between 0 and the
sliding limit $\sigma_t=\mu \sigma_n$ (in a 2D case)
\cite{Radjai1996}. We use these different observations to infer
the form of the "rejuvenated" distribution $P_{rej}(\sigma_n,
\sigma_t)$, which is thus written:

\begin{equation}
P_{rej}(\sigma_n, \sigma_t) = \frac{1}{\sigma_0 \cdot \sigma_n}
\cdot \exp \left( {- \frac{\sigma_n }{ \sigma_0}} \right)
\end{equation}

\noindent where $\sigma_0$ is the mean stress inside the
pile. For simplicity, we have omitted the plateau
saturation of the distribution at low forces. We can now derive
the dynamical equation of evolution of the stress distribution
$P_{\sigma}$:

\begin{eqnarray}
  \frac{\partial P_{\sigma}(\sigma_n, \sigma_t)}{\partial t}&=& -P_{\sigma}(\sigma_n, \sigma_t) \cdot \omega_y(\sigma_n, \sigma_t) \\
  &&\lfloor   +    \; \; P_{rej}(\sigma_n, \sigma_t) \cdot F(P_{\sigma}) \nonumber
\label{dyn_Psigma}
\end{eqnarray}

\noindent where $F(P_{\sigma})$ is the total frequency
of sliding events which is self-consistently defined as:
\begin{equation}
F(P_{\sigma}) = \int \!\!\!\! \int_{\sigma'_t < \sigma'_n}
\!\!\!\! P_{\sigma}(\sigma'_n, \sigma'_t)\cdot \omega_y(\sigma'_n,
\sigma'_t) \cdot d\sigma'_n d\sigma'_t \label{freqslidevent}
\end{equation}

The present description exhibits many common features with
Bouchaud's trap model \cite{Bouchaud1992} of glass transition. In
the latter, the internal dynamics of a glassy liquid is pictured
as a succession of thermal escapes from energy wells of various
depths. In an analogous way, each contact here can be considered
as frozen in a mechanical trap (the local solid friction cone),
the depth of which depends on the relative amplitude of the normal
and shear components of the contact force. Moreover, in the
absence of temperature, mechanical vibrations plays the role of
the energy source by allowing individual contacts to hop out of
their trap.

\begin{figure}[!h]
\centerline{\epsfxsize=8.5truecm \epsfbox{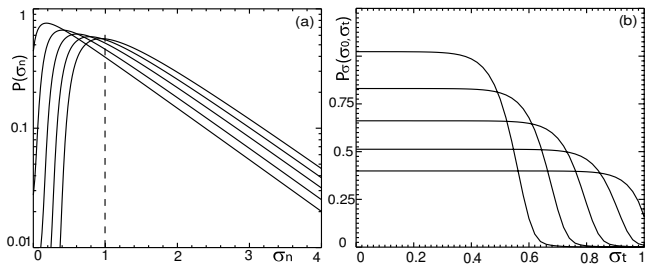}}
\caption{Results of the numerical model for $\sigma_0 = 1$ and
$\overline{\delta\sigma}=\sigma_0/20$: (a) Distributions of normal
forces for $t_w = 1,10,100,1000$ and $10000$ (from left to right);
(b) distributions of tangential forces for $\sigma_n = \sigma_0$
at the same times $t_w$ (from bottom to top).}
 \label{figure4}
 \end{figure}

As in the trap model, we thus observe two limiting regimes
depending on the relative values of the intensity of the applied
stress $\overline {\delta\sigma}$ and the width $\sigma_0$ of the
rejuvenated distribution. For large vibrations, {\it i.e.}
$\overline {\delta\sigma}>\sigma_0$, the rejuvenated stress
distribution is a stationary solution of Eq. (\ref{dyn_Psigma}),
and the yielding frequency is constant with time. By contrast, for
$\overline {\delta\sigma}<\sigma_0$, the stress distribution keeps
evolving endlessly. Figure \ref{figure4} shows the time evolution
of $P_{\sigma}$ obtained by numerically solving Eq.
(\ref{dyn_Psigma}) for $\overline{\delta\sigma}=\sigma_0/20$,
starting with $P_{\sigma}=P_{rej}$. It shows that fragile contacts
- contacts of low normal force or close to the sliding limit
(inset) - are slowly depleted. As a result, the number of sliding
events per time unit decays. More quantitatively, we find that the
characteristic time between events grows linearly with the elapsed
time. This is to be compared with the $\tau \sim t_w^{1.2}$
scaling behavior observed in the experiment (Figure
\ref{figure3}.)

We have evidenced, through MSDWS measurements, the existence of a
slowing down of the micro-scale dynamics over more than five
decades in gently vibrated granular piles. This behavior is
reminiscent of the aging process observed in glassy
systems. This dynamics appears to be weakly connected to the overall
grain-scale structure, which suggests a two-level description of
granular systems. At high enough vibration, a granular pile
evolves through the restructuration of the piling geometry,
leading to a slow irreversible compaction. In this regime, forces
networks are rapidly renewed and show no history-dependent
behavior. At very low vibrations however, the geometry of the pile
is essentially frozen, but the forces network can still evolve by
slowly depleting the most fragile contacts. This leads to an
effective reinforcement of the pack structure as is evidenced in the
present study by the decrease of vibration induced plastic events.

The precise nature of the yielding events remains however unclear.
In particular, one might expect large spatial and temporal
correlations between them, which we cannot probe with DWS. Another
important question concerns the relevance of such modifications to the
onset of macroscopic flow. For
instance, does the observed reinforcement of the force networks
play a role in changing the threshold of avalanche triggering or
shear-banding appearance ?

\begin{acknowledgments}

We wish to thank Luca Cipelletti and Jean-Marc di Meglio for
fruitful discussions and Deniz Gunes for helping us with the
estimation of $l^{\star}$.

\end{acknowledgments}


\begin{thebibliography}
\bibitem[{\citenamefont{note-l-star}(2000)}]{note-l-star}

$l^\star$ was found equal to $220 \pm 20 \mu m$, which allows in
our geometry the use of a diffusive model for the light in the
sample \cite{DWS}.
\end{thebibliography}
\end{document}